# Coherent oscillations in a single electron spin at room temperature

F. Jelezko, T. Gaebel, I. Popa, A. Gruber, J. Wrachtrup
3. Physikalisches Institut, Universität Stuttgart, Germany

**Abstract**

Rabi nutations of a single electron spin in a single defect center have been detected. The coherent evolution of the spin quantum state is followed via optical detection of the spin state. Coherence times up to several microseconds at room temperature have been measured. Optical readout of the spin states leads to decoherence, equivalent to the so-called Zeno effect. Quantum beats between electron spin transitions in a single spin Hahn echo experiment are observed. A closer analysis reveals that beats also result from the hyperfine coupling of the electron spin to a single $^{14}$N nuclear spin. The results are analysed in terms of a density matrix approach of an electron spin interacting with two oscillating fields.

Detection and manipulation of single electron spin states in solids have recently received much attention in the context of quantum computing [1-4]. Mainly, this is because localized electron spins in solids show long relaxation [5] and coherence times and their states can be easily manipulated via microwave or radio frequency pulses. Yet reading out a single spin state, a prerequisite for most quantum algorithms, has remained a major challenge up to now. Although single spin detection has been reported previously [6] [7], most of the systems and techniques have not shown the ability to measure the quantum state of a single electron spin. The only solid state system where this has been successful is the nitrogen-vacancy (NV) defect center in diamond. The defect is known to have a strongly dipole allowed optical transition between its electron spin triplet ground state ($^3$A) and a first excited spin triplet state ($^3$E) (Fig.1) [8]. Fluorescence emission strongly depends on the electron spin quantum state, which has lead to optically detected magnetic resonance on single defect centers [9]. Electron spin relaxation times ($T_1$) range from 2ms at room temperature to seconds at low temperature. Recently, electron spin quantum jumps have been detected [10], thus showing that the spin state of a single electron spin could be determined in this defect center. In this Letter we report on the generation of coherence among single electron spin states in the NV center. We analyse how spin coherence is influenced by readout and coupling to additional degrees of freedom, like single $^{14}$N nuclear spins. Single spin coherence in solids has been observed before [11, 12]. However, the NV center in diamond is particularly interesting because of its electron paramagnetic ground state. Contrary to other systems, like organic molecules, spin coherence time is thus not limited by the electronic lifetime of the state. Above all, the electron spin state can be readout directly via optical excitation and subsequent fluorescence relaxation. Previously this has been the basis of a number of ingenious studies on NV center ensembles, like the observation of electromagnetic induced transparency [13] and dressed state nutation [14].

Experiments have been carried out with a home-build confocal microscope operating in a temperature range between 2K and 300K. Microwaves are coupled to the sample by a miniaturized short cut loop connected to a 40W travelling wave tube amplifier to generate microwave H fields large enough for short microwave pulses. As sample material, synthetic nanocrystalline diamond ( Type 1b) with an average crystal size of 20nm was used. To allow for comparison with a density matrix based theory, all experiments have been time averaged over $10^5$ to $10^7$ cycles to obtain smooth curves.

The ground and excited electron states of the NV defect are spin triplets (S=1) with spin sublevels $m_s$=0,±1 (Fig. 1) [15]. The degeneracy among the sublevels is lifted due to either the mutual dipolar interaction of the two unpaired electron spins or an external magnetic field. Most of the experiments in this paper have been performed without applying an external magnetic field. Because of the $C_{3v}$ symmetry of the defect, the $m_s$=±1 sublevels remain degenerated (E=0) but are split from the $m_s$=0 sublevel by roughly 2.9 GHz. Fluorescence of a single defect center is visible only when the spin is in the $m_s$=0 spin sublevel. Optical excitation leads to a strong spin polarization, such that, averaged in time, the electron spin is found with at least 70% probability in the $m_s$=0 sublevel. The application of a resonant microwave pulse causes a transition of the system between the spin levels and thus modulates the fluorescence intensity. When time averaged, the nutation of the electron spin due to coherent interaction of the spin with the microwave H-field is described as

$r_3(t) = r_3(0)\frac{1}{\varpi^2}(\Delta\omega^2 + \omega^2_1 \cos\varpi t)$, where $\Delta\omega$ is the detuning between microwave and transition frequency and $\varpi = \sqrt{\Delta\omega^2 + \omega_1^2}$, with $\omega_1$ as microwave Rabi frequency [16]. $r_3(t)$ is the difference between the probability to find the system in the $m_s$=0 and $m_s$=±1 sublevel. Figure 2 shows the nutation of a single NV center electron spin. Since coupling to the microwave and optical field has to be considered, the above mentioned equation has to be extended and experiments are compared with microwave-optical Bloch equations starting from the semiclassical Hamiltonian

$$H = \sum_i \hbar \omega_i |i\rangle\langle i| - \hbar\Omega\cos(\omega_L t)(|1\rangle\langle 3| + |3\rangle\langle 1|) - \hbar\omega_1 \cos(\omega_M t)(|1\rangle\langle 2| + |2\rangle\langle 1|).$$

In this equation $\hbar\omega_i$ is the energy of the ground and excited states, $\Omega = E_0 d_{13}/\hbar$ is the Rabi frequency corresponding to the amplitude of the laser electric field $E_0$, at frequency $\nu_L = \omega_L/2\pi$, interacting with the defect center transition dipole moment $d_{13}$. $\omega_1 = \gamma H_1$ is the Rabi frequency of the ESR transition between the $m_s = 0$ and $\pm 1$ states with frequency $\nu_M = \omega_M/2\pi$. The interaction between the molecule and the applied field is described by the Liouville equation for the electronic density matrix $\rho$, $i\hbar\dot{\rho} = [H, \rho]$, in which all the populations are considered. However, all coherences, non-resonant with an external field may be neglected. Weak coupling to a broad spectrum of bath states (phonons and surrounding spins) causes electronic dephasing and relaxation. After averaging over the bath states, the coupling appears in the reduced Liouville equation as dephasing and relaxation rate constants. The resulting set of differential equations is solved numerically, with the dephasing time of the electron spin as a fit parameter. The decay measured in the nutation experiment is determined by the electron spin dephasing time $T_2$, which in our experiments ranges from 1.5 to 2 μs, depending on the defect center under study. Recently, ensemble experiments have shown that $T_2$ in this system can reach values larger than 30 μs [13], with increasing $T_2$ in samples with low nitrogen content. Our samples were not optimised for low nitrogen concentration, which explains the relatively short $T_2$ times in the present experiments. The most important source for dephasing in this system occurs via spin flip-flop processes, either directly between electron spins of the NV center and residual nitrogen impurities in the diamond lattices (P-center, S=1/2) or via hyperfine coupling to the nitrogen nuclear spin. In both cases, the dephasing rate is strongly distance dependent ($1/r^3$). We found a substantial difference among the $T_2$ values of different

NV centers, possibly due to a change in the distance between the electron spin of the center and the next nearest neighbour nitrogen in the lattice.

In the experiment, we observed a decrease in the decay time of the spin nutation upon increased optical excitation (readout) intensity. Figure 3 shows experimental data together with theoretically predicted curves. In the microwave optical Bloch equations, the differential equation for the coherence between the spin sublevels $\rho_{12}$ does not directly comprise optical pumping. Rather the effect of optical pumping enters via $\rho_{11}$, which is a function of the optical Rabi frequency. As a consequence, the time constant for fluorescence recovery after a microwave pulse shows the same dependence on optical pumping as the coherence. Assuming a single spin experiment that is not time averaged, our finding is explained as follows. A measurement leads to a projection of the single spin in one of its eigenstates. Consequently, an increase in the measurement frequency is followed by a decrease in damping time of spin coherence. Similar experiments have been reported for single ions and have been related to the Zeno effect [17], although there has been a considerable debate on this subject. We note that our experimental results are accurately described by Bloch equations, which do not explicitly encounter state collapse. Neither in experiment nor in theory we found a linear relation between nutation damping and optical Rabi frequency. Rather the damping time follows the saturation of the fluorescence intensity of the defect. In the microwave optical Bloch equations spin decoherence caused by optical excitation is governed by the dependence of $\rho_{11}$ on the optical Rabi frequency. It thus follows the same saturation behaviour as the populations. Assuming non time-averaged measurement on a single spin, these results may indicate that the spontaneous fluorescence decay (or the subsequent detection of the fluorescence photon) causes spin state projection and not the optical excitation itself. An important prerequisite for probing unperturbed electron spin coherence is thus to carry out spin manipulation in the absence of optical excitation. In this scheme, state preparation results from optical excitation, i.e. the spin is brought to the $m_s=0$ state via optical pumping.

Subsequently, microwave pulses are applied and the final result is measured via laser excitation and fluorescence emission. The experiments described below follow this scheme. In order to probe how well the single spin state can be coherently manipulated by microwave pulses, the spin state evolution under a Hahn echo pulse sequence has been investigated. Adapted to optically detected magnetic resonance, the Hahn echo pulse sequence consists of the well-known 90°-τ-180°-τ´ microwave pulses and waiting times followed by a 90° pulse to convert the spin echo phenomenon into populations, measurable by fluorescence detection [18]. Figure 4 shows the single spin Hahn echo together with echo decay measurements. The Hahn echo sequence refocuses inhomogeneously distributed ESR transition frequencies, if this distribution is static on the time scale of the echo sequence. Indeed, when time averaged, we found an inhomogeneously broadened ESR resonance line. The source of the inhomogeneity becomes apparent when the Hahn echo amplitude is followed as a function of the waiting time τ=τ´. Superimposed on the echo decay which marks $T_2$, a strong modulation shows up. This echo envelope modulation is a well known phenomenon from bulk ESR measurements [19]. It results from a beating between different transitions within the ESR resonance spectrum, all excited coherently by the microwave pulse. A Fourier transform of the echo decay function after subtracting the exponential decay itself, reveals a characteristic frequency of 17 MHz. This corresponds well with the measured splitting between the $m_s$=+1 and −1 state of the particular defect studied. This splitting is the result of a slight deviation from $C_{3v}$ symmetry of the defect. Since the microwave Rabi frequency $\omega_1$ is substantially larger than the frequency splitting $\Delta\omega$ between the two states, both transitions are excited simultaneously. To confirm our interpretation a small external magnetic field $H_0$ was applied. The Hamiltonian which now describes the system is $\boldsymbol{H}= \boldsymbol{S}\boldsymbol{D}\boldsymbol{S}+g_e\beta_e\boldsymbol{S}\boldsymbol{H}_0+\boldsymbol{S}\boldsymbol{A}\boldsymbol{I}-g_n\beta_n\boldsymbol{I}\boldsymbol{H}_0$, where $\boldsymbol{D}$ and $\boldsymbol{A}$ are the fine structure and hyperfine splitting (hfs) tensors, and $g_{e,n}$ and $\beta_{e,n}$ are the electron and nuclear g-factors and Bohr magnetons, respectively. The electron Zeeman term further splits the $m_s=\pm1$ levels. For an $H_0$ field of 0.02T, the splitting is roughly 150 MHz, such that the microwave pulse only excites the $m_s=0$ to $-1$ transition. As a result the beating in the Hahn echo due to the interference between the two electronic transitions disappears. Instead, a modulation with much lower frequency appears. The Fourier transform now shows frequency components between 5 and 10 MHz. Following a previous analysis of Hahn echo decay data on NV center ensembles [20], we believe that this envelope modulation arises from hyperfine coupling of the electron spin to $^{14}$N nuclear spins. Due to the low spin density of the electrons at the location of the nucleus, the hyperfine coupling of $^{14}$N usually is not resolved in the ESR spectrum [21]. Although the frequency components caused by $^{14}$N are also visible in the echo decay data for $H_0 = 0$ T, they are considerably more pronounced when a small magnetic field is applied. In order to detect the nuclear modulation effect allowed ($\Delta m_I=0$) as well as forbidden ($\Delta m_I=\pm1$), ESR transitions have to be excited. The modulation depth is largest when the nuclear Zeeman splitting $\omega_I$ equals the hyperfine frequency, i.e., $\omega_I=|m_s|A_{iso}$, where $A_{iso}$ is the isotropic hyperfine coupling. Under this condition, the nuclear spin is no longer quantized along the external magnetic field. The allowed and forbidden ESR transitions have same magnitude, and the modulation depth is at maximum [19]. This condition is met for the experiment shown (Fig. 4b) where $H_0=0.02$T. Hyperfine coupling to $^{14}$N also explains the beating seen in Fig.4a. If $\omega_M$ is detuned by $\Delta\omega$ from the spin resonance transition, the echo is expected to be modulated by $\cos(\Delta\omega(\tau-\tau'))$, where $\Delta\omega$ is the hyperfine coupling. Indeed, the observed modulation frequencies correspond well with those found in the Fourier transform of the Hahn echo decay in an external magnetic field.

In summary, we report on the first experimental investigation of coherent oscillations among spin sublevels of a ground state single electron spin at room temperature. The measured dephasing times range from 1µs to 2µs. They strongly depend on the optical readout frequency, which is in good agreement with microwave optical Bloch equations. The coherence times are long enough to allow for a coherent manipulation of the single spin. These experiments reveal the hyperfine interaction of the electron with the $^{14}$N nuclear spin of the defect center. Since our experiments demonstrate a coherent evolution of electron spin transitions for different nitrogen nuclear spin configurations, they may be a first step towards conditional non-trivial two single spin quantum logical operations.

**Figure captions**

Fig. 1. Energy level scheme of the nitrogen vacancy defect center in diamond. The excited state spin manifold can be approximated by a single spin sublevel, since non-resonant optical excitation has been used in the experiment (see text). The greyed out lines correspond to the $m_s = \pm 1$ sublevels.

Fig. 2. Optically detected Rabi oscillations of single electron spins in a single NV defect center are shown for two values of the MW Rabi frequency; (a) corresponds to a slow MW Rabi frequency (aprox. 16 MHz), and (b) corresponds to a faster Rabi frequency (39 MHz) . The solid grey lines represent the measured data, and the thick lines show the simulations based on the microwave optical Bloch equations described in the text. The inset shows the dependence of the observed modulation frequency on the applied microwave magnetic field.

Fig. 3. Damping of spin coherence as a function of optical Rabi frequency. The MW power was held constant, while the laser power was gradually increased. (a) and (b) represent the nutation curves for two different values of the laser power (corresponding laser Rabi frequencies were 40 MHz and 7 MHz, respectively). The experimental data are shown in grey, and the theoretical curves in bold lines. The inset shows the nutation damping time versus optical Rabi frequency. The symbols represent the experimental data. The curve was calculated according to the model presented in the text. The nutation damping time is decreasing upon increasing the optical excitation power.

Fig. 4 (a) Hahn echo of a single NV center electron spin. The applied MW sequence is depicted together with the experimental data. The Hahn echo occurs for a value of the delay time τ' = τ = 0.3 μs. (b) Hahn echo decay curve of a single electron spin at zero external field and in a field $H_0$=0.02 T. The inset shows the Fourier transform of the respective echo decays after subtraction of the exponential decay term (the dashed line).

References

[1] B. E. Kane, Nature **393**, 133 (1998).

[2] Clark R.G., B. R. and B. T. M. PHILOS T ROY SOC A **361**, 1451-1471 (2003).

[3] A. M. Stoneham, A. J. Fisher and P. T. Greenland, J Phys Condens Matter **15**, L447-L451 (2003).

[4] Ed. D. D. Awschalom, D. Loss and N. Samarth, Semiconductor Spintronics and Quantum Computation, Springer Verlag, Berlin 2002.

[5] G. Feher, Phys Rev **114**, 1219-1244 (1959).

[6] J. Köhler, J. A. J. M. Disselhorst, M. C. J. M. Donckers, E. J. J. Groenen, J. Schmidt and W. E. Moerner, Nature **363**, 242-244 (1993).

[7] J. Wrachtrup, C. von Borczyskowski, J. Bernard, M. Orrit and R. Brown, Nature **363**, 244-245 (1993).

[8] Ed. G. Davies, Properties and growth of Diamond, INSPEC, London 1994.

[9] A. Gruber, A. Dräbenstedt, C. Tietz, L. Fleury, J. Wrachtrup and C. von Borczyskowski, Science **276**, 2012-2014 (1997).

[10] F. Jelezko, I. Popa, A. Gruber and J. Wrachtrup, Appl Phys Lett **81**, 2160-2162 (2002).

[11] J. Wrachtrup, C. von Borczyskowski, J. Bernard, M. Orrit and R. Brown, Phys Rev Lett **71**, 3565-3568 (1993).

[12] A. C. J. Brouwer, E. J. J. Groenen and J. Schmidt, Phys Rev Lett **80**, 3944-3947 (1998).

[13] E. A. Wilson, N. B. Manson and C. Wei, Phys Rev A **67**, 023812-1 023812-10 (2003).

[14] C. Wei, N. B. Manson and J. P. D. Martin, Phys Rev Lett **74**, 1083-1086 (1995).

[15] D. A. Redman, S. Brown, R. H. Sands and S. C. Rand, Phys Rev Lett **67**, 3420-3423 (1991).

[16] W. G. Breiland, H. C. Brenner and C. B. Harris, J Chem Phys **62**, 3458 (1975).

[17] W. M. Itano, J. J. Bollinger and D. J. Wineland, Phys Rev A **41**, 2295-2300 (1989).

[18] W. G. Breiland, C. B. Harris and A. Pines, Phys Rev Lett **30**, 158 (1973).

[19]  W. B. Mims, Phys Rev B **5**, 2409-1419 (1972).

[20]  F. T. Charnock and T. A. Kennedy, Phys Rev B **64**, 041201-1-041201-4 (2001).

[21]  X.-F. He, N. B. Manson and P. T. H. Fisk, Phys Rev A **47**, 8816-8822 (1992).

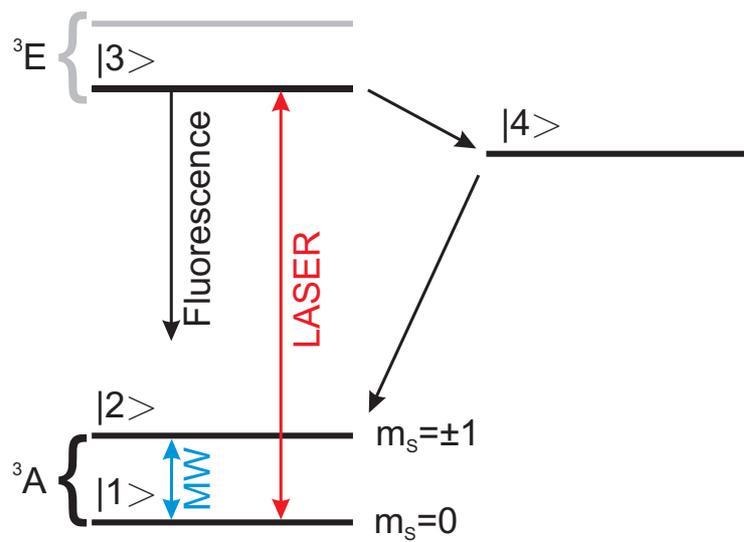

Figure 1

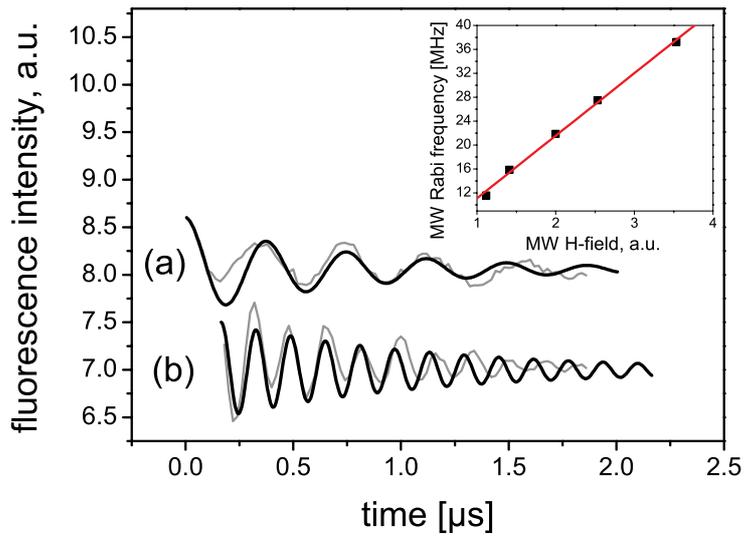

Figure 2

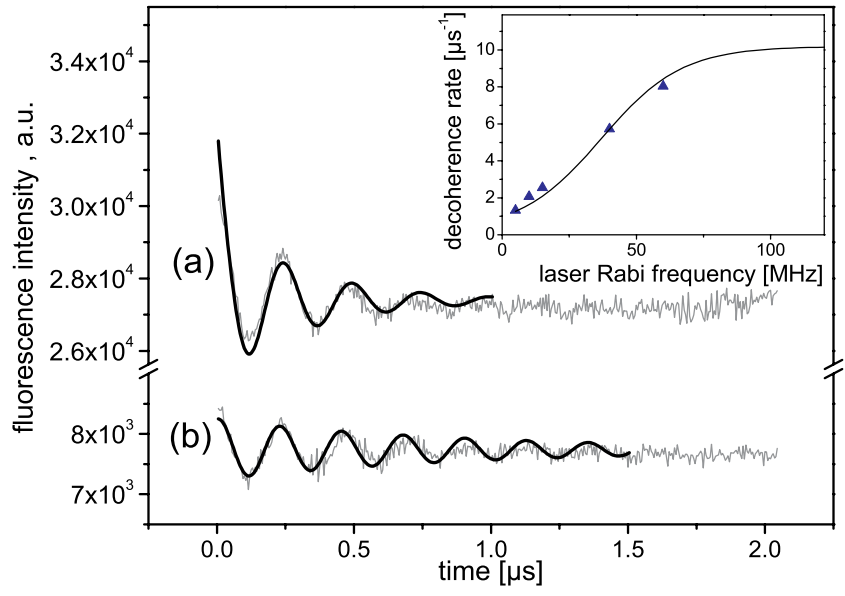

Figure 3

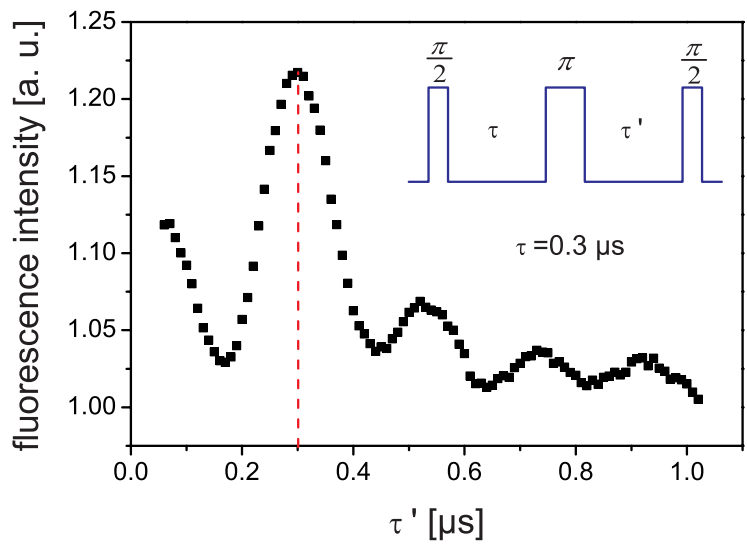

Figure 4a

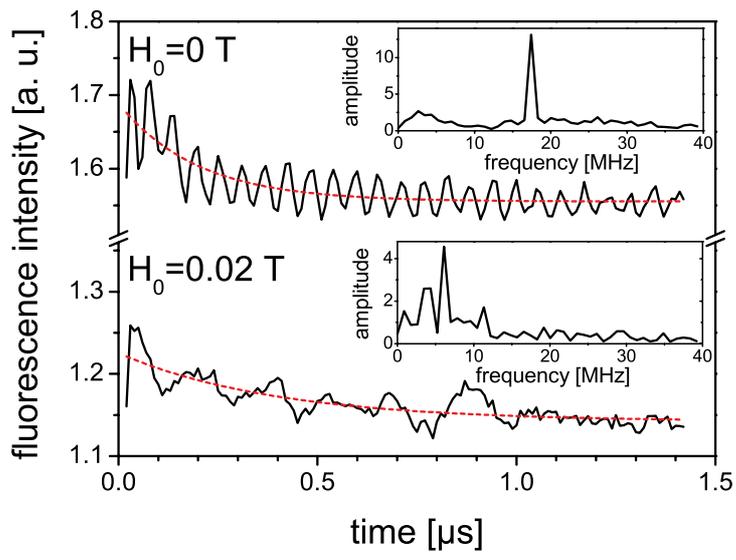

Figure 4b